\documentclass[10pt,conference,letterpaper]{IEEEtran}
\IEEEoverridecommandlockouts

\usepackage{cite}
\usepackage{amsmath,amssymb,amsfonts}
\usepackage{graphicx}
\graphicspath{{picture/}}
\usepackage{textcomp}
\usepackage{xcolor}
\usepackage{booktabs}
\usepackage{multirow}
\usepackage{tabularx}
\usepackage{hyperref}
\hypersetup{hidelinks}

\def\BibTeX{{\rm B\kern-.05em{\sc i\kern-.025em b}\kern-.08em
    T\kern-.1667em\lower.7ex\hbox{E}\kern-.125emX}}

\begin{document}

\title{Scaling Adaptive Non-Local Observable Quantum Super-Resolution via Matrix Product States}

\author{
\IEEEauthorblockN{
    Shih-Lung Yu\IEEEauthorrefmark{1},
    Ming-Kang Ho\IEEEauthorrefmark{2},
    Tai-Yue Li\IEEEauthorrefmark{2},
    and Sheng Yun Wu\IEEEauthorrefmark{1}
}
\IEEEauthorblockA{\IEEEauthorrefmark{1} Department of Physics, National Dong Hwa University, Hualien, Taiwan}
\IEEEauthorblockA{\IEEEauthorrefmark{2} National Center for High-performance Computing, National Institutes of Applied Research, Hsinchu, Taiwan}
\IEEEauthorblockA{
Emails: 811214201@gms.ndhu.edu.tw, 2603013@niar.org.tw, tim312508@gmail.com, sywu@mail.ndhu.edu.tw}
}

\maketitle

\begin{abstract}
This work presents a matrix product state (MPS) simulation framework for adaptive non-local observable variational quantum circuits (ANO-VQCs) in image super-resolution (SR) beyond the practical limits of statevector simulation. Runtime benchmarks on a single NVIDIA RTX 4070 GPU show that, under the tested shallow-circuit setting, MPS completes ANO-VQC forward feature extraction for individual inputs up to 16 x 16 pixels (256 qubits), whereas statevector simulation encounters a memory bottleneck at 6 x 6 inputs (36 qubits) and exact tensor-network (Exact TN) contraction becomes computationally impractical beyond 12 x 12 inputs (144 qubits). For a fixed 7 x 7 input (49 qubits), a bond-dimension sweep over depths L = 1 to L = 4 shows that the required MPS bond dimension increases with circuit depth. Using Exact TN contraction as the reference, the bond dimension required for near-exact agreement increases from chi = 2 at L = 1 to chi = 16 at L = 4. Finally, 7 x 7 to 28 x 28 Fashion-MNIST SR training with chi = 16 shows that the shallow L = 1 model achieves the lowest loss, lowest LPIPS, and highest PSNR and SSIM among the tested depths. These results highlight MPS as a scalable and controllable simulation backend for ANO-VQC image SR and as a practical tool for studying large-scale quantum algorithms.
\end{abstract}

\begin{IEEEkeywords}
Matrix Product States, Tensor Networks, Adaptive Observables, Variational Quantum Circuits, Image Super-resolution.
\end{IEEEkeywords}

\section{Introduction}
\label{sec:introduction}

Image super-resolution (SR) reconstructs a high-resolution image from a low-resolution input and is commonly addressed using classical deep-learning models~\cite{dong2016srcnn}. In parallel, variational quantum algorithms have been explored as an alternative computational framework~\cite{cerezo2021variational}. Related studies have applied quantum kernels and VQC-based neural modules to classification, convolution, channel attention, and image representation~\cite{tai2022quantum,an2025quantum,hsu2025qae,wang2026mpmqir}. In VQCs, classical data are encoded into quantum states, processed by quantum gates, and converted into task-relevant features through expectation values.

Conventional VQCs usually extract features from fixed Pauli observables. Adaptive non-local observable variational quantum circuits (ANO-VQCs)~\cite{lin2026quantum} instead treat Hermitian measurement operators as trainable components, enabling task-dependent combinations of multiqubit correlations. For image SR, these adaptive-observable expectation values form a quantum feature vector that is mapped to the high-resolution output by a classical reconstruction layer.

A central scalability bottleneck is statevector simulation. With one-pixel-to-one-qubit encoding, a $p \times p$ input requires $n=p^2$ qubits and a statevector with $2^n$ complex amplitudes. Tensor networks provide compact representations for such high-dimensional states and have been used to simulate large-scale quantum machine-learning and circuit workloads~\cite{vidal2003efficient,orus2014practical,chen2024validating,sam2026dual,sam2026grover}. In particular, a matrix product state (MPS) represents an $n$-qubit state as a chain of local tensors connected by virtual bonds. The bond dimension $\chi$ controls the retained correlations, and the storage requirement scales approximately as $\mathcal{O}(n\chi^2)$ for fixed $\chi$, rather than $\mathcal{O}(2^n)$.

This study proposes an MPS simulation framework for ANO-VQC image SR. The framework maps each low-resolution pixel to one qubit, restricts trainable two-qubit Hermitian observables to image-local neighborhoods, and uses Pauli-basis decomposition to cache fixed Pauli expectation values during training. This design separates expensive quantum-feature evaluation from the trainable observable and reconstruction layers, making larger-qubit ANO-VQC simulation more practical. We evaluate three aspects: runtime scalability across statevector simulation, exact tensor-network (Exact TN) contraction, and MPS simulation; the MPS accuracy-cost trade-off over circuit depths and bond dimensions for 49-qubit ANO circuits; and the resulting workflow on a $7\times7\rightarrow28\times28$ Fashion-MNIST SR task.

\section{Methodology}
\label{sec:method}

\begin{figure*}[!t]
    \centering
    \includegraphics[width=0.96\textwidth]{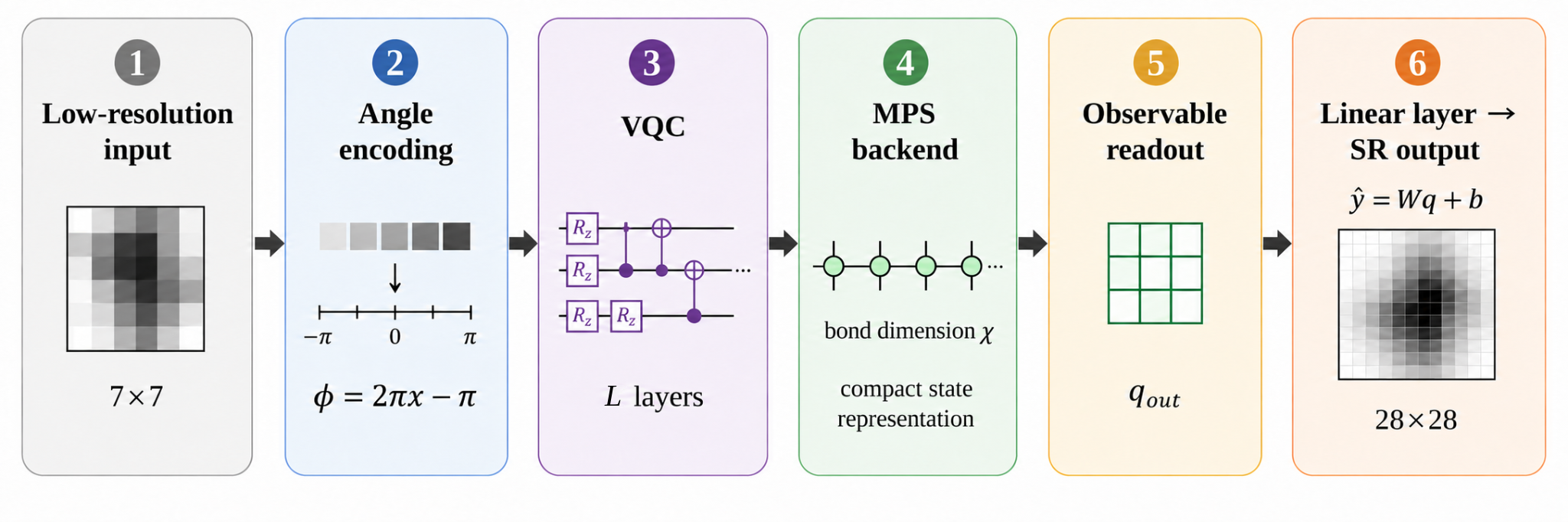}
    \caption{Workflow of the proposed MPS-ANO-VQC framework for
    Fashion-MNIST SR. A $7 \times 7$ low-resolution image is mapped to a
    49-qubit input state and simulated with an MPS backend. Image-local
    adaptive observables produce quantum features, which are passed to a
    classical linear layer to reconstruct the $28 \times 28$ high-resolution
    image.}
    \label{fig:mps_ano_vqc_architecture}
\end{figure*}

\subsection{Adaptive Non-Local Observable Variational Quantum Circuit}
ANO-VQC uses trainable Hermitian observables instead of fixed Pauli measurements. In this framework, the quantum circuit is input-dependent, while the adaptive observables and the classical reconstruction layer contain the trainable parameters. Each observable acts on a qubit pair $e_q=(i_q,j_q)$ and is represented locally by a $4 \times 4$ Hermitian matrix $H_q$, constructed from real-valued parameter sets $A_q$, $B_q$, and $D_q$ for the real off-diagonal, imaginary off-diagonal, and diagonal components. For expectation evaluation, $H_q$ is embedded into the full Hilbert space as $H_q^{(e_q)}$, and the extracted feature is
\begin{equation}
q_q=\left\langle\psi(\mathbf{x})\middle|H_q^{(e_q)}\middle|\psi(\mathbf{x})\right\rangle
\end{equation}
where $\lvert \psi(\mathbf{x}) \rangle$ denotes the quantum state generated from the low-resolution image $\mathbf{x}$. The expectation values form the quantum feature vector $\mathbf{q}=[q_1,q_2,\ldots,q_m]^{\mathrm{T}}$, where $m$ denotes the number of image-local two-qubit observables.

\subsection{Fashion-MNIST Encoding and MPS Simulation}
Fashion-MNIST images at $28 \times 28$ pixels are used as high-resolution targets, and $7 \times 7$ resized images are used as low-resolution inputs. Under one-pixel-to-one-qubit encoding, the 49 input pixels are flattened and mapped to 49 qubits. Each normalized pixel value $x_i \in [0,1]$ is converted into a rotation angle
\begin{equation}
\phi_i = 2\pi x_i - \pi
\end{equation}
The encoding circuit applies a Hadamard gate to every qubit, followed by pixel-dependent $R_Y(\phi_i)$ rotations and staggered nearest-neighbor CNOT gates. The circuit has no independently trainable gate angles because the $R_Y(\phi_i)$ rotations are determined directly by the input image. In the depth studies, $L$ denotes the number of repeated encoding-entangling layers.

A 49-qubit statevector contains $2^{49}$ complex amplitudes. To avoid storing this vector explicitly, the quantum state is represented as a matrix product state (MPS)
\begin{equation}
\lvert \psi \rangle=\sum_{s_1,\ldots,s_{49}}A^{[1]s_1}A^{[2]s_2}\cdots A^{[49]s_{49}}\lvert s_1 s_2 \cdots s_{49} \rangle
\end{equation}
where $s_i \in \{0,1\}$ is the physical index of the $i$-th qubit, and $A^{[i]s_i}$ is the corresponding local tensor. The maximum virtual dimension is the bond dimension $\chi$, which controls the retained correlations. For fixed $\chi$, storage scales approximately as $\mathcal{O}(n\chi^2)$ instead of $\mathcal{O}(2^n)$. One-qubit gates update local tensors, whereas nearest-neighbor two-qubit gates update adjacent tensors followed by bond-dimension truncation. The scalability benchmark uses $L=1$ and $\chi=8$, the bond-dimension sweep uses $\chi=2$--2048, and the SR training experiment uses $\chi=16$.

\subsection{Quantum Feature Extraction and Image Reconstruction}
To preserve image locality and reduce measurement cost, two-qubit observables are assigned to horizontal, vertical, and diagonal neighbors on the $7 \times 7$ image grid. This gives 156 neighboring qubit pairs instead of 1,176 all-to-all pairs. To evaluate the adaptive Hermitian observables through fixed Pauli measurements, each observable is expanded in the Pauli basis as 
\begin{equation} 
H_q = \sum_{\alpha} c_{q,\alpha} P_{\alpha} 
\end{equation} 
where $P_{\alpha}\in\{I,X,Y,Z\}^{\otimes 2}$ is embedded on $e_q$ when evaluating its expectation value, with $c_{q,\alpha}=2^{-k}\operatorname{Tr}(H_qP_{\alpha})$ and $k=2$. Using the linearity of expectation values, the output of the $q$-th adaptive observable is computed as 
\begin{equation} 
\langle H_q \rangle = \sum_{\alpha} c_{q,\alpha} \langle P_{\alpha} \rangle 
\end{equation} 
Thus, the MPS backend evaluates only fixed Pauli expectation values, with trainable coefficients determined by $A_q$, $B_q$, and $D_q$. Since the encoding circuit depends only on the input image, these Pauli expectation values are computed once and cached. Training then updates the observable coefficients and reconstruction layer without repeatedly executing the MPS circuit for the same image. The resulting ANO feature vector is passed through a trainable linear reconstruction layer:
\begin{equation} 
\hat{\mathbf{y}} = W\mathbf{q}+\mathbf{b}
\end{equation} 
where $W$ and $\mathbf{b}$ denote the trainable weight matrix and bias vector, respectively. The output vector contains $28^2=784$ values and is subsequently reshaped into the predicted $28\times 28$ high-resolution Fashion-MNIST image.

\subsection{Model Training and Optimization}
The reconstructed image $\hat{\mathbf{y}}$ is compared with the corresponding $28 \times 28$ ground-truth image $\mathbf{y}$ using a weighted combination of the mean squared error (MSE) and the learned perceptual image patch similarity (LPIPS) loss:
\begin{equation}
\mathcal{L}=\lambda_{\mathrm{MSE}}\mathcal{L}_{\mathrm{MSE}}+\lambda_{\mathrm{LPIPS}}\mathcal{L}_{\mathrm{LPIPS}}
\end{equation}
where the weighting coefficients are set to 
$\lambda_{\mathrm{MSE}}=0.3$ and $\lambda_{\mathrm{LPIPS}}=0.7$.
During backpropagation, separate Adam optimizers are used to update the Hermitian-observable parameters 
$A$, $B$, and $D$, and the linear-layer parameters $W$ and $\mathbf{b}$.

The reported loss denotes the weighted MSE--LPIPS objective, and LPIPS is also reported separately. Reconstruction quality is evaluated using peak signal-to-noise ratio (PSNR) and structural similarity index measure (SSIM), with PSNR computed on the image-intensity scale.

All runtime and training experiments were conducted on a single NVIDIA RTX 4070 GPU. The Fashion-MNIST experiment used 10{,}000 training images, 200 test images, and a batch size of 8.

\section{Results}
\label{sec:results}

\begin{figure}[!t]
    \centering
    \includegraphics[width=0.98\columnwidth]{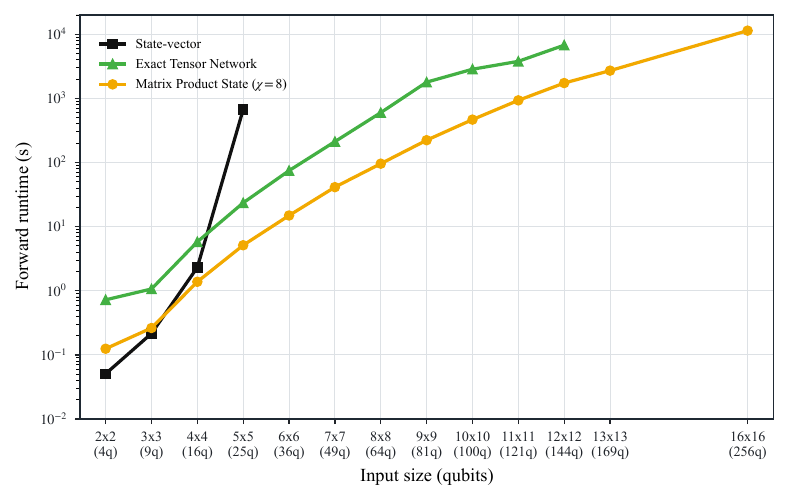}
    \caption{Forward-runtime comparison among statevector simulation, Exact TN
    contraction, and MPS simulation for ANO-VQC quantum-feature extraction as
    the input image size and qubit count increase. The scalability benchmark
    uses $L=1$, and MPS uses a fixed bond dimension of $\chi=8$.}
    \label{fig:runtime_backend_comparison}
\end{figure}

The experiments evaluate backend runtime scalability, the MPS accuracy-cost trade-off at $7\times7$ inputs (49 qubits), and Fashion-MNIST SR performance under the selected near-exact setting. Figure~\ref{fig:runtime_backend_comparison} compares statevector simulation, Exact TN contraction, and MPS simulation with $\chi=8$ for input sizes from $2\times2$ to $16\times16$ (4 to 256 qubits). Runtime includes circuit execution and Pauli expectation-value evaluation. Statevector simulation is fastest for very small inputs, but reaches a memory bottleneck at $6\times6$ inputs (36 qubits) after requiring approximately 671~s at $5\times5$. Here, memory bottleneck denotes out-of-memory allocation, and impractical denotes exceeding the preset benchmark runtime budget. Exact TN contraction remains executable up to $12\times12$ inputs (144 qubits), but becomes impractical at $13\times13$ inputs (169 qubits). In contrast, MPS completes the tested forward feature-extraction runs through $16\times16$ inputs (256 qubits), with a runtime of approximately 11{,}399~s. Over the overlapping input-size range, MPS is faster than Exact TN contraction and becomes faster than statevector simulation at approximately $4\times4$ inputs (16 qubits).

\begin{figure}[!t]
    \centering
    \includegraphics[width=0.98\columnwidth]{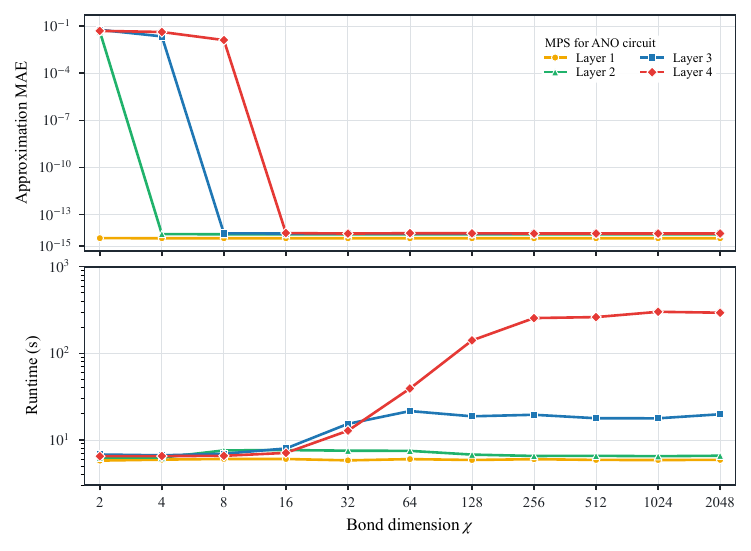}
    \caption{Influence of the MPS bond dimension $\chi$ on approximation
    accuracy and runtime for 49-qubit ANO circuits with $L=1$--4 layers.
    Top: mean absolute error (MAE) between MPS and the Exact TN
    contraction reference. Bottom: MPS simulation runtime as a function of $\chi$. The $\chi$
    axis is shown on a log$_2$ scale.}
    \label{fig:bond_dimension_tradeoff}
\end{figure}

The second experiment evaluates the bond dimension required for accurate MPS feature extraction at the target $7\times7$ input size. The circuit depth is varied from $L=1$ to $L=4$, and the MPS bond dimension is swept from $\chi=2$ to 2048. Each MPS result is compared against Exact TN contraction using the mean absolute error (MAE),
\begin{equation}
\operatorname{MAE}(\chi)=\frac{1}{M}\sum_{m=1}^{M}\left|q_{\mathrm{MPS},m}(\chi)-q_{\mathrm{ExactTN},m}
\right|
\label{eq:mps_mae}
\end{equation}
where $M$ is the number of compared Pauli expectation values, and $q_{\mathrm{MPS},m}(\chi)$ and $q_{\mathrm{ExactTN},m}$ denote the $m$-th values from MPS and Exact TN, respectively. Near-exact agreement is defined as $\operatorname{MAE}\leq10^{-12}$. Figure~\ref{fig:bond_dimension_tradeoff} shows that the required bond dimension increases with circuit depth. The first near-exact points occur at $\chi=2$, 4, 8, and 16 for $L=1$, 2, 3, and 4, with runtimes of approximately 5.78~s, 6.26~s, 6.93~s, and 7.09~s, respectively. For $L=4$, the MAE decreases from $4.91\times10^{-2}$ at $\chi=2$ to $1.27\times10^{-2}$ at $\chi=8$ and $6.83\times10^{-15}$ at $\chi=16$. Increasing $\chi$ beyond 16 keeps the $L=4$ MAE near $10^{-15}$ but increases runtime to 12.79~s, 39.32~s, and 141.19~s at $\chi=32$, 64, and 128, respectively. Thus, $\chi=16$ is used in the final experiment as the smallest tested bond dimension that reaches near-exact agreement for all depths from $L=1$ to $L=4$.

\begin{figure}[!t]
    \centering
    \includegraphics[width=\columnwidth]{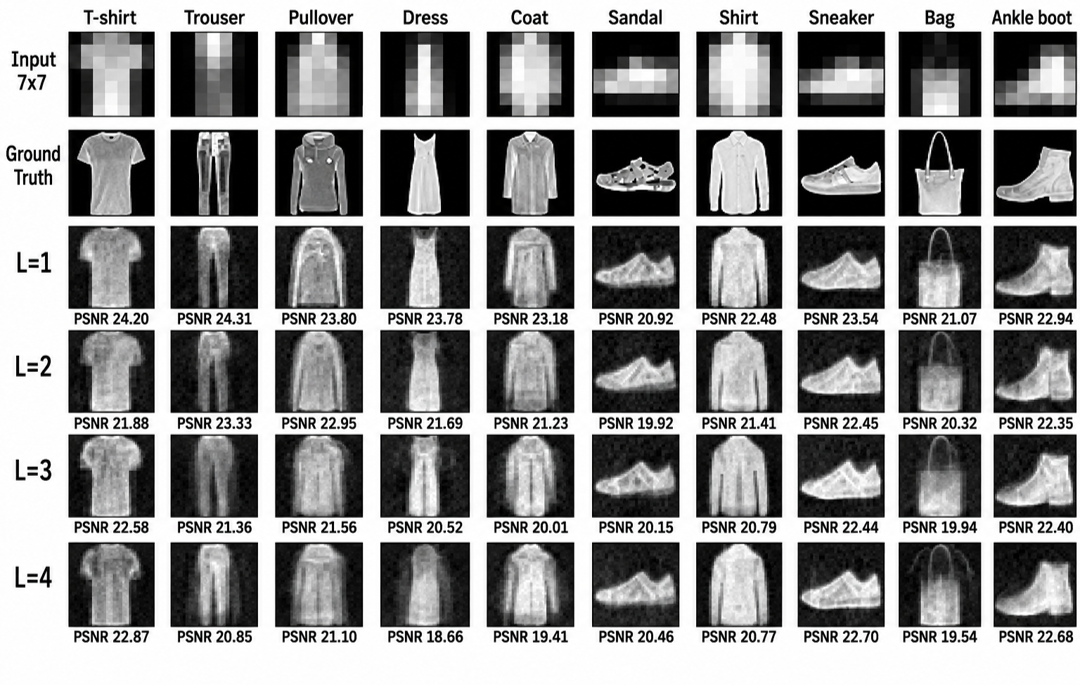}
    \caption{Representative Fashion-MNIST SR reconstructions for different
    circuit depths using a fixed MPS bond dimension of $\chi=16$.}
    \label{fig:fashion_mnist_depth_results}
\end{figure}

With $\chi=16$, the third experiment trains the image-local adaptive observables and the classical reconstruction layer for $7\times7\rightarrow28\times28$ Fashion-MNIST SR~\cite{xiao2017fashion}. Table~\ref{tab:depth_ablation} reports the aggregate depth-ablation metrics, while Fig.~\ref{fig:fashion_mnist_depth_results} shows representative reconstructions. The shallow $L=1$ circuit gives the lowest loss, lowest LPIPS, and highest PSNR and SSIM among the tested depths. Relative to $L=4$, $L=1$ lowers loss by 30.4\% and LPIPS by 12.3\%, increases PSNR by 2.07~dB and SSIM by 45.4\%, and reduces training time by 0.81~h. The representative examples are consistent with the quantitative trend, with $L=1$ preserving clearer object boundaries than deeper circuits. Thus, increasing circuit depth does not improve average reconstruction quality under this fixed near-exact MPS setting.

\begin{table}[!t]
\centering
\caption{Depth ablation for $7\times7\rightarrow28\times28$ Fashion-MNIST SR using MPS with $\chi=16$. Loss denotes the weighted objective $0.3\,\mathrm{MSE}+0.7\,\mathrm{LPIPS}$; PSNR is computed on the image-intensity scale.}
\label{tab:depth_ablation}
\footnotesize
\setlength{\tabcolsep}{6.5pt}
\begin{tabular}{@{}cccccc@{}}
\toprule
Layers & Time (h) & Loss & LPIPS & PSNR & SSIM \\
\midrule
1 & \textbf{3.20} & \textbf{0.4864} & \textbf{0.3194} & \textbf{20.87} & \textbf{0.4248} \\
2 & 3.40 & 0.5835 & 0.3448 & 19.79 & 0.3541 \\
3 & 3.89 & 0.6400 & 0.3540 & 19.26 & 0.3206 \\
4 & 4.01 & 0.6984 & 0.3640 & 18.80 & 0.2922 \\
\bottomrule
\end{tabular}
\end{table}

\section{Conclusion}
In the present single-GPU benchmarks, MPS completes ANO-VQC forward feature extraction for individual inputs through $16\times16$ pixels (256 qubits), extending the tested forward-runtime range beyond the statevector and Exact TN backends. For $7\times7$ inputs (49 qubits), deeper ANO circuits require larger bond dimensions to reach near-exact agreement; at $\chi=16$, all tested depths from $L=1$ to $L=4$ achieve MAE values on the order of $10^{-15}$. In $7\times7\rightarrow28\times28$ Fashion-MNIST SR, the shallow $L=1$ circuit achieves the lowest loss, lowest LPIPS, and highest PSNR and SSIM among the tested depths. These results suggest that circuit depth must be selected jointly with the observable design and reconstruction layer, since deeper circuits do not necessarily improve SR quality under a fixed near-exact MPS setting. The same backend can also support reinforcement-learning studies in which circuit topology or observable layout is optimized under explicit accuracy-runtime rewards. Overall, MPS provides a practical tool for studying large-scale variational quantum circuits. Future work will combine reinforcement-learning-based circuit search, parallelization, circuit cutting, and distributed tensor-network computation to further extend the simulation scale and investigate larger-scale ANO SR algorithms.

\section*{Acknowledgment}
The authors acknowledge the National Center for High-performance Computing (NCHC), National Institutes of Applied Research (NIAR), Taiwan, for computational resources and research infrastructure. OpenAI ChatGPT was used for language editing; the authors reviewed all AI-assisted text and take responsibility for the final content.

\bibliographystyle{IEEEtran}
\bibliography{reference}

\end{document}